\documentclass[12pt,a4]{article}%a5paper
\usepackage[T1]{fontenc}
\usepackage{fixmath}
\usepackage{amsmath}
\usepackage{amssymb}
\usepackage{natbib}
\usepackage{graphicx}
\usepackage{subfigure}
\usepackage{xcolor}
\usepackage{multirow}

\def\er{\mathbb{R}}
\def\bx{{\mathbf x}}
\def\bX{{\mathbf X}}

\def\beq{\begin{eqnarray*}}
\def\eeq{\end{eqnarray*}}

\title{Nonparametric Regression and Classification with Functional, Categorical, and Mixed Covariates}
\author{Leonie Selk $^1$ \& Jan Gertheiss $^{1,2}$\\\\
{\small $^1$ Helmut-Schmidt-University, Department of Mathematics and Statistics,}\\{\small Hamburg, Germany,} \\{\small $^2$ Department of Biostatistics, Epidemiology, and Informatics,}\\{\small Perelman School of Medicine,}\\{\small University of Pennsylvania, Philadelphia, USA}\\\\{\small corresponding author: leonie.selk@hsu-hh.de}}

\begin{document}
	
\maketitle

\begin{abstract}
We consider nonparametric prediction with multiple covariates, in particular categorical or functional predictors, or a mixture of both. The method proposed bases on an extension of the Nadaraya-Watson estimator where a kernel function is applied on a linear combination of distance measures each calculated on single covariates, with weights being estimated from the training data. The dependent variable can be categorical (binary or multi-class) or continuous, thus we consider both classification and regression problems. The methodology presented is illustrated and evaluated on artificial and real world data. Particularly it is observed that prediction accuracy can be increased, and irrelevant, noise variables can be identified/removed by `downgrading' the corresponding distance measures in a completely data-driven way.  

%For prediction models with multiple categorical and/or functional predictors we consider a nonparametric estimation procedure to predict the scalar, dependent variable that can be categorical (binary or multi-class) or continuous, thus we consider classification as well as regression models. Besides the prediction we focus on variable selection to distinguish between relevant predictors and noise variables.
%Our method bases on an extension of the Nadaraya-Watson estimator where we apply a kernel function on a sum of weighted distances. These weights are the crucial terms in our model and we estimate them via cross-validation. The relative sizes of the estimated weights then indicate how relevant the covariates are for the prediction.
%\\
%The paper describes the methodology of our procedure and shows its applicability. We present an extensive simulation study and apply our method to several real world data examples including trajectories of participants movements in a psychological experiment.

{\bf Classification} 62G08 \and 62H12 \and 62H30 \and 62P15
\end{abstract}

\section{Introduction}\label{introduction}

We consider nonparametric prediction and estimation with multiple categorical or functional predictors, or a mixture of both. Especially in the case of a categorical, multi-class response, the number of corresponding methods found in the literature is very limited.
%We focus on categorical and functional predictors and a mixture of both with $p>1$ since in these cases there are very few alternative methods available for nonparametric prediction, especially in the case of a categorical multi-class response. 

The proposed method is an expansion of the well-known Nadaraya-Watson estimator 
\[\hat{f}(x) = \frac{\sum_{i=1}^{n} Y_i K((X_i-x)/h_n)}{\sum_{i=1}^{n} K((X_i-x)/h_n)},\] 
with some kernel $K(\cdot)$ and bandwidth $h_n\searrow 0$ (for $n \rightarrow \infty$),
that was introduced by \cite{Nadaraya1964} and \cite{Watson1964} as a nonparametric estimator for the regression function in a model $Y_i=f(X_i)+\varepsilon_i$ with continuous observations $(X_1,Y_1),\ldots,(X_n,Y_n)$. In the classification case with categorical response $Y$ this estimator can be adapted to estimate the posterior probability $P_g(x)=P(Y=g|x)$ as
\[\hat{P}_g(x) = \frac{\sum_{i=1}^{n} I\{Y_i=g\} K((X_i-x)/h_n)}{\sum_{i=1}^{n} K((X_i-x)/h_n)},\] 
see for instance \cite{HasTibFri2009}. We extend these estimators to handle multiple functional, categorical or mixed predictors, see Section \ref{Methods}. Besides estimation of the regression function we are interested in variable selection, thus in separating relevant predictors from noise variables. For this sake we determine some weights (counterpart to bandwidth) for each covariate in a data-driven way, see Section \ref{Methods} for details. The size of the weights then indicates the relevance of the corresponding covariate. For a recent review on variable selection for regression models with functional covariates particularly see \cite{AneirosNovoVieu2022}.

Existing methods for nonparametric classification/regression and variable selection as covered by the method proposed in the paper at hand can be arranged in four macro-areas by the type of response (categorical/continuous) and predictor (functional/categorical). The case of a categorical response and functional predictors is handled, for instance, in \cite{FuchsGertheissTutz2015} who use an ensemble approach for classification of multiple functional covariates. They estimate the posterior probability separately for every covariate and weight the results to get an estimate of the overall posterior probability. Further they use several semi-metrics and combine the results in an analogous way. Thus their method can be used for feature as well as variable selection. 
A similar approach is followed by \cite{GulEtal2018} for categorical responses and categorical or continuous covariates. They use an ensemble of kNN classifiers based on random subsets of the covariates with the aim to select the most relevant covariates.
The same type of response-covariate combination is considered by \cite{MbinaNkietObiang2019} who propose a procedure for classification in more than two groups with  categorical (binary) and continuous predictors. Their aim is to select among the continuous variables those that are relevant for the classification. They use a criterion to quantify the loss of information resulting from selecting not all continuous variables and compare different procedures to estimate the criterion's value.
Continuous responses are considered e.\,g.\ in  \cite{Shang2014} and \cite{RacineHartLi2006}.  
\cite{Shang2014} considers a nonparametric regression model with a mixture of functional, categorical and continuous covariates. He uses a Bayesian approach to determine simultaneously the different bandwidths. His method can also be used for variable selection since the irrelevant variables are smoothed out by the appropriate bandwidth.
\cite{RacineHartLi2006} test for significance of categorical predictors in regression models with categorical and continuous predictors. They use a product kernel to estimate the regression function and approximate the distribution of their test statistic under the null using a bootstrap procedure.

\begin{figure}
%\begin{centering}
\includegraphics[width=0.33\textwidth]{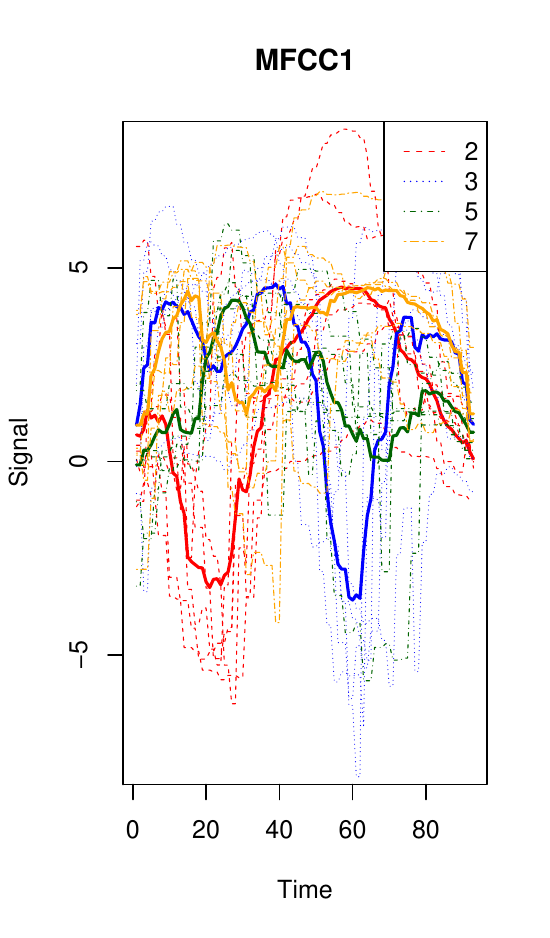}
\hspace{-2mm}\includegraphics[width=0.33\textwidth]{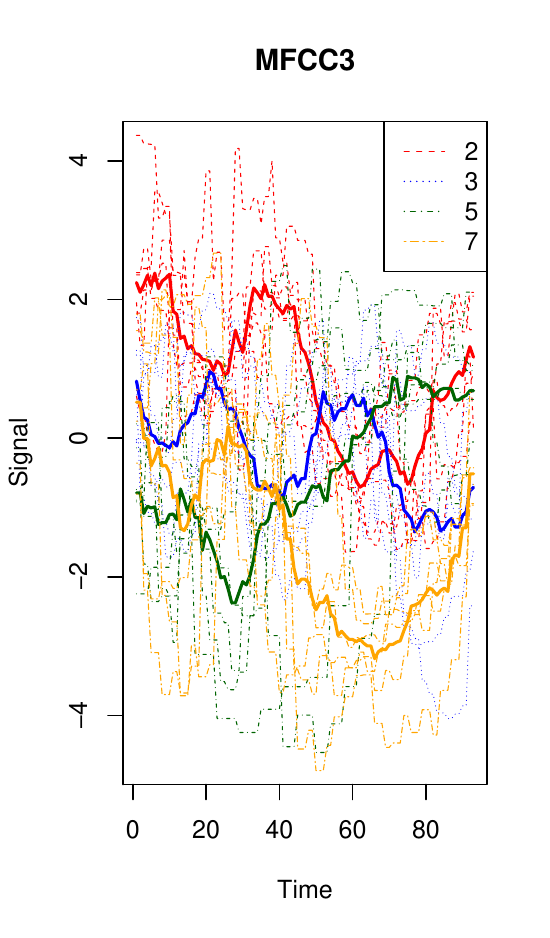}
\hspace{-2mm}\includegraphics[width=0.33\textwidth]{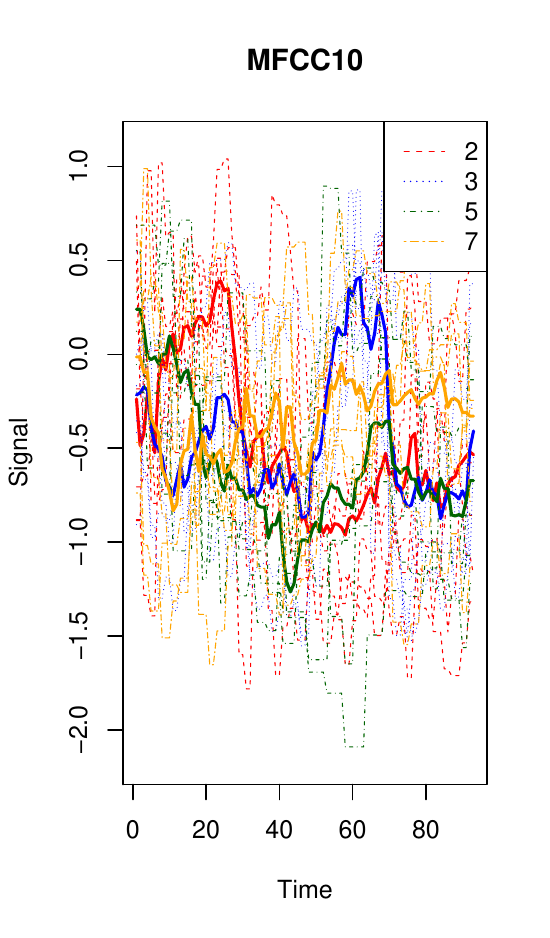}
%\end{centering}
\caption{Illustration of the ArabicDigits in terms of a subset of the available signals for three Mel Frequency Cepstrum Coefficients, and digits `2', `3', `5', and `7'; solid lines correspond to the respective mean curves.}\label{fig:mfcc}
\end{figure}

As an application example of the procedure proposed here, consider the following classification problem: the well-known {\it ArabicDigits} data set from the R-package {\it mfds} by \cite{Rmfds}, which contains time series of 13 Mel Frequency Cepstrum Coefficients (MFCCs) corresponding to spoken Arabic digits. MFCCs are very common for speech recognition, see \cite{KoolagudiRastogiRao2012} for a detailed explanation. Figure \ref{fig:mfcc} shows a subset of the available signals for MFCC1, MFCC3 and MFCC10, and digits  `2', `3', `5', and `7'. In total, and more generally speaking, we are faced with a 10-class problem (digits $0,1,\ldots,9$) and 13 functional predictors.

The rest of the paper is organized as follows. In Section~\ref{Methods}, we begin with the regression case to explain the idea of our approach, and then put our focus on classification problems. Both cases are investigated through simulation studies in Section~\ref{NumExp}. The real data mentioned above and some further data, such as trajectory data from a psychological, virtual reality experiment, is revisited in Section~\ref{RealData}, illustrating the presented method's broad spectrum of potential applications. Section~\ref{Conclude} concludes with a short discussion and outlook.

\section{Methodology}\label{Methods}

Suppose there are training data $\bX_i=(X_{i1},\ldots,X_{ip})$, $i=1,\ldots,n$, with variables contained in $\bX_i$ being continuous, categorical, functional, or a mixture of those. In addition, there is information $Y_i$ on a scalar, dependent variable which may be continuous or categorical. 

\subsection{Regression}\label{Remod}

Let us first consider the regression problem with continuous $Y_i$ and a single covariate $X_i$, where
\[Y_i = f(X_i) + \varepsilon_i,\] 
$f$ being an unknown regression function, and $\varepsilon_i$ some mean zero noise variable, potentially with some further assumptions such as independent identically distributed (iid) across subjects $i = 1,\ldots,n$.\\

For a new observation with known covariate value $x$, but unknown $Y$, a kernel-based, nonparametric prediction $\hat{Y} = \hat{f}(x)$ is, e.g., given by 
\[\hat{f}(x) = \frac{\sum_{i=1}^{n} Y_i K(d(X_i,x)/h_n)}{\sum_{i=1}^{n} K(d(X_i,x)/h_n)},\] 
with some kernel $K(\cdot)$, bandwidth $h_n\searrow 0$ (for $n \rightarrow \infty$) and distance measure $d(\cdot,\cdot)$ that is appropriate for the type of predictor considered. In particular with functional data, $d(\cdot,\cdot)$ may also be calculated through so-called \emph{semi}-metrics, compare \cite{FerratyVieu} and Section~\ref{Metric} below.\\

Now suppose for multiple (and potentially very different) predictors as given above, there are $d_1(\cdot,\cdot),\ldots,d_p(\cdot,\cdot)$ available. With categorical predictors $X_{il},x_l \in \{1,\ldots,G_l\}$, for example, we may use
\begin{equation}\label{distcat}d_l(X_{il},x_l) = \left\{ \begin{array}{ll}
0 & \mbox{ if } X_{il} = x_l,\\
1 & \mbox{ if } X_{il} \neq x_l,
\end{array}
\right.\end{equation}
or for functional $X_{ij},x_j \in L^2$, for instance,
\begin{equation}\label{metric}
d_j(X_{ij},x_j) = \sqrt{\int_{\mathcal{D}_j} (X_{ij}(t) - x_j(t))^2 dt},
\end{equation}
where $\mathcal{D}_j$ is the domain of the functions $X_{ij},x_j$. In what follows, we will omit the $\mathcal{D}_j$ for the sake of readability.

\bigskip

When predicting $Y$, multivariate predictor information $\bx=(x_1,\ldots,x_p)$ should be considered jointly. A somewhat natural way to do so, appears to be
\begin{equation}\label{fhat}\hat{Y}=\hat{f}(\bx) = \frac{\sum_{i=1}^{n} Y_i K(\omega_1d_1(X_{i1},x_1) + \ldots + \omega_pd_p(X_{ip},x_p))}{\sum_{i=1}^{n} K(\omega_1d_1(X_{i1},x_1) + \ldots + \omega_pd_p(X_{ip},x_p))},\end{equation}
with positive weights $\omega_1,\ldots,\omega_p$ that should be estimated from the data. With $\hat{Y}_{(-i)}$ being the leaving-one-out estimate
\[\hat{Y}_{(-i)} = \frac{\sum_{s\neq i} Y_s K(\omega_1d_1(X_{s1},X_{i1}) + \ldots + \omega_pd_p(X_{sp},X_{ip}))}{\sum_{s\neq i} K(\omega_1d_1(X_{s1},X_{i1}) + \ldots + \omega_pd_p(X_{sp},X_{ip}))},\]
we may estimate $\omega_1,\ldots,\omega_p$ by minimizing
\begin{equation}\label{Qre}
Q(\omega_1,\ldots,\omega_p) = \sum_{i=1}^n (Y_i - \hat{Y}_{(-i)})^2.
\end{equation}
By $\hat\omega_1,\ldots,\hat\omega_p$ we denote these minimizing weights.

The nonparametric estimator $\hat f$ defined in \eqref{fhat} is an extension of the well known Nadaraya-Watson estimator, see Section \ref{introduction}. Similar extensions of this kind of kernel estimator to the multivariate case are also well established; see, e.g., \cite{HaerdleMueller1997} for some deeper insight. The typical form of a multivariate Nadaraya-Watson estimator for continuous covariates is 
\[\hat f_{\text{NW}1}(\bx)=\frac{\sum_{i=1}^nY_iK(|X_{i1}-x_1|/h_1)\cdot\ldots\cdot K(|X_{ip}-x_p|/h_p)}{\sum_{i=1}^nK(|X_{i1}-x_1|/h_1)\cdot\ldots\cdot K(|X_{ip}-x_p|/h_p)}\]
with bandwidths $(h_1,\ldots,h_p)$, or
\[\hat f_{\text{NW}2}(\bx)=\frac{\sum_{i=1}^nY_iK(\|{\mathbf H}^{-1}(\bX_i-\bx)\|)}{\sum_{i=1}^nK(\|{\mathbf H}^{-1}(\bX_i-\bx)\|)}\]
where $\|\cdot\|$ is, e.g., the euclidean norm and ${\mathbf H}$  is a symmetric bandwidth matrix. If we set $K$ in our $\hat f$ defined in \eqref{fhat} as an exponential function, e.\,g., the Picard kernel $K(u)=e^{-u}I\{u\geq 0\}$, we have a very similar setting to $\hat f_{\text{NW}1}$ with $|X-x|$ replaced by the more general $d(X,x)$. Also, our $\hat f$ can be interpreted as a form of $\hat f_{\text{NW}2}$ with $\|\cdot\|$ being some kind of $L_1$-norm (Manhattan-norm). Estimation of the weights (\ref{Qre}) is similar to determining an optimal bandwidth for the Nadaraya-Watson estimator with cross-validation. There are different possibilities to choose the starting values for the numerical minimization of $Q(\omega_1,\ldots,\omega_p)$. In the simulation studies to follow in Section~\ref{NumExp}, for instance, we will use a rule of thumb for the bandwidth size for the regression case, whereas for the classification case (see Section~\ref{ClMod} below) we determine a pre-estimator for each weight by considering $p$ models each with only one predictor.

\subsection{Classification}\label{ClMod}

In the classification case, we also consider models that may contain functional and/or categorical predictors $X_{ij}$ ($i=1,\ldots,n$, $j=1,\ldots,p$), but categorical responses $Y_i \in\{1,\ldots,G\}$ for $i=1,\ldots,n$ instead of continuous ones.
Especially the case $p>1$, $G>2$ is of interest since in this `multi$^2$fun' case (multiple, possibly functional predictors and a multi-class response) there are only very few genuinely nonparametric methods available (compare Section \ref{introduction}).

Following the idea for the regression case, we estimate the posterior probability $P_g(\bx):=P(Y=g|\bx)$
for a new set of predictor values $\bx=(x_1,\ldots,x_p)$ with unknown class label $Y$ by
\[\hat P_g(\bx)=\frac{\sum_{i=1}^nI\{Y_i=g\}K\left(\omega_1d_1(X_{i1},x_1)+\ldots+\omega_pd_p(X_{ip},x_p)\right)}{\sum_{i=1}^nK\left(\omega_1d_1(X_{i1},x_1)+\ldots+\omega_pd_p(X_{ip},x_p)\right)}\]
with data-driven weights $\omega_1,\ldots,\omega_p$. As before we determine the weights by minimizing
\begin{equation}\label{Qcla}
Q(\omega_1,\ldots,\omega_p)= \sum_{i=1}^n\sum_{g=1}^G (I\{Y_i=g\} - \hat{P}_{g(-i)})^2
\end{equation}
where $\hat P_{g(-i)}$ is the leave-one-out estimator
\[\hat P_{g(-i)}=\frac{\sum_{s\neq i}I\{Y_s=g\}K\left(\omega_1d_1(X_{s1},X_{i1})+\ldots+\omega_pd_p(X_{sp},X_{ip})\right)}{\sum_{s\neq i}K\left(\omega_1d_1(X_{s1},X_{i1})+\ldots+\omega_pd_p(X_{sp},X_{ip})\right)}.\]
Quantity \eqref{Qcla} is also know as the Brier score or quadratic scoring rule, compare \cite{GneRaf2007}, \cite{Brier1950} and \cite{Selten1998}.

\subsection{Distances and (Semi-)Metrics}\label{Metric}

A crucial question when dealing with functional predictors is the choice of the (semi-)metric $d$, contrary to models with predictors that take values in $\er^p$, since in a finite dimensional euclidean space all norms are equivalent. This concept fails for functional predictors since they take values in an infinite dimensional space. Even more, restricting $d$ to be a metric is sometimes too restrictive in the functional case. That is why semi-metrics are considered such as
\begin{equation}\label{semi}
d(u,v) = \sqrt{\int (u'(t) - v'(t))^2 dt},\end{equation}
where $u,v$ are functional predictors and $u',v'$ their derivatives, see \cite{FerratyVieu} Chapter 3 for a deeper insight on this topic. An important difference between semi-metric \eqref{semi}, for instance, and a metric is that in the former case $d(u,v) = 0$ will also be obtained if $v(t) = u(t) + c$, for some constant $c \neq 0$, i.\,e., if $v$ is just a vertically shifted version of $u$. In general, the choice which (semi-)metric to take depends on the shape of the data and the goal of the statistical analysis. If, for example, the functional observations shall be displayed in a low-dimensional space, one possibility to do this is to use (functional) principal component analysis; compare, e.g., \cite{RamsaySilverman} and \cite{YaoMueWan2005}. In general, results can look very different, depending on the chosen measure of proximity. In  Chapter~3 of \cite{FerratyVieu} examples to illustrate this effect are given. Also, further suggestions for semi-metrics and a survey which semi-metric may be appropriate for which situation can be found there. For example, semi-metric~\eqref{semi}, which is based on the derivatives, is often well suited for smooth data whereas for rough data a different approach should be considered.

The (semi-)metric also plays an important role for the asymptotic properties of nonparametric functional estimators. Chapter 13 in  \cite{FerratyVieu} is dedicated to this issue. The small ball probability that is defined as $P(d(u,v)<\epsilon)$ appears in the rate of convergence of many nonparametric estimators such as the functional Nadaraya-Watson estimator. If the small ball probability decays very fast when $\epsilon$ tends to zero (in other words, if the functional data are very dispersed) the rate of convergence will be poor, whereas a small ball probability decaying adequately slowly will lead to a rate of convergence similar to those found in finite dimensional settings.

In our simulation studies and for the real data examples we will use a form of the $L^2$-metric as already given in~\eqref{metric}.
%\[d(x_1,x_2) = \sqrt{\int (x_1(t) - x_2(t))^2 dt}\]
%for functional predictors $x_1,x_2$.
This is a standard choice which works quite well for our examples. Note that, although our focus in this paper is not on the choice of the distance measure, our procedure could also be used to give a data driven answer on the question which (semi-)metric to choose. For this sake let us suppose there is only one functional predictor with observations $X_1,\ldots,X_n$ and a set of $p$ potential (semi-)metrics $d_1,\ldots,d_p$. With this we set 
$$\hat f(x)=\frac{\sum_{i=1}^nY_iK(\omega_1d_1(X_i,x)+\ldots+\omega_pd_p(X_i,x))}{\sum_{i=1}^nK(\omega_1d_1(X_i,x)+\ldots+\omega_pd_p(X_i,x))}$$
and
$$\hat P_g(x)=\frac{\sum_{i=1}^nI\{Y_i=g\}K(\omega_1d_1(X_i,x)+\ldots+\omega_pd_p(X_i,x))}{\sum_{i=1}^nK(\omega_1d_1(X_i,x)+\ldots+\omega_pd_p(X_i,x))},$$
respectively. Then, the estimated weights $\hat\omega_1,\ldots,\hat\omega_p$ tell us which distance measures are appropriate to explain the influence of the covariate on the response: those that are weighted highest. This approach is especially useful for feature selection since the (semi-)metrics can be chosen such that each $d_j$ focuses on a certain feature of the curve, compare to \cite{FuchsGertheissTutz2015}.

\section{Numerical Experiments}\label{NumExp}

\subsection{Regression Problems}\label{ReSim}

\subsubsection{Set-up}
To investigate the finite sample performance of our procedure, we generate data according to a model with mixed covariates (MixR), combining functional and categorical predictors. For $i=1,\ldots,n$, we generate functional covariates $X_{i1},\ldots,X_{ip_{\text{fun}}}$ according to
$$\tilde X_{ij}(t)=\sum_{l=1}^5\left(B_{ij,l}\sin\left(\frac tT(5-B_{ij,l})2\pi\right)-M_{ij,l}\right),$$ 
where $B_{ij,l}\sim\mathcal{U}[0,5]$ and $M_{ij,l}\sim\mathcal{U}[0,2\pi]$ for $l=1,\ldots,5$, $j=1,\ldots,p$, $i=1,\ldots,n$, and $T=300$.  $\mathcal{U}$ stands for the (continuous) uniform distribution. Then, $X_{ij}(t)$ is calculated from $\tilde X_{ij}(t)$ by scaling it in direction $i$ and then dividing each value by 10. 
The categorical covariates are generated as
$X_{i(p_{\text{fun}}+1)},\ldots,X_{i(p_{\text{fun}}+p_{\text{cat}})}$ $\sim${\it B}$(0.5)$, such that $p_{\text{fun}}+p_{\text{cat}}=p$. With this we get an extended functional linear model
$$Y_i=5\sum_{j=1}^{q_{\text{fun}}}\int X_{ij}(t)\gamma_{3,\frac 13}(t/10)dt+2(X_{i(p_{\text{fun}}+1)}+\ldots+X_{i(p_{\text{fun}}+q_{\text{cat}})})+\varepsilon_i$$
for some $q_{\text{fun}}\leq p_{\text{fun}}$ and $q_{\text{cat}}\leq p_{\text{cat}}$, where the coefficient function $\gamma_{a,b}(t)=b^a/\Gamma(a)t^{a-1}e^{-bt}I\{t>0\}$ is the density of the Gamma distribution. See \cite{RamsaySilverman} Chapter 15 or \cite{KokoszkaReimherr} Chapter 4 for an introduction to functional linear models. The errors $\varepsilon_i$ are iid standard normal. Further simulation examples (FunR, CatR) with solely functional or categorical covariates can be found in the online supplement.

We investigate `minimal' and `sparse' cases. %(where `sparse' refers to the number of influential predictors, not the type of functional data) 
Specifically,  we compare the cases  $q_{\text{fun}}=q_{\text{cat}}=1$, $p_{\text{fun}}=p_{\text{cat}}=2$ (minimal: (*.m)) and $q_{\text{fun}}=q_{\text{cat}}=2$, $p_{\text{fun}}=p_{\text{cat}}=8$ (sparse: (*.s)). For all generated data sets we use a one-sided Picard kernel $K(u)=e^{-u}I\{u\geq 0\}$ and the results shown are based on 500 replications each.

To uncouple the estimation of the weights from the bandwidth that goes to zero as $n$ grows, we set 
\begin{eqnarray*}
d_{\text{fun}}(X_{ij},x_j)&:=&\frac 1{h_n^\text{fun}c_j^\text{fun}}\sqrt{\int(X_{ij}(t)-x_j(t))^2dt},\\
d_{\text{cat}}(X_{ij},x_j)&:=&\frac 1{h_n^\text{cat}c_j^\text{cat}}\sqrt{(X_{ij}-x_j)^2}\\
&=&\frac 1{h_n^\text{cat}c_j^\text{cat}}I\{X_{ij}\neq x_j\},
\end{eqnarray*}
with norming constants 
$$c_j^\text{fun}=\sqrt{\int \frac1{n-1}\sum_{l=1}^n\left(X_{lj}(t)-\frac 1n\sum_{k=1}^nX_{kj}(t)\right)^2dt},$$ 
$$c_j^\text{cat}=\sqrt{\frac1{n-1}\sum_{l=1}^n\left(X_{lj}-\frac 1n\sum_{k=1}^nX_{kj}\right)^2},$$  
and bandwidths $h_n^\text{fun}=n^{-\frac1{p+4}}$ and $h_n^\text{cat}=\frac{p+4}{\ln(n)}$, respectively. This choice of bandwidths coincides with the order of the optimal bandwidths in \cite{RacineLi2004} when $K$ is the one sided Picard kernel and the categorical covariates are {\it B}$(0.5)$-distributed.

The prediction is then calculated as given in~\eqref{fhat}. For MixR, this means
\[\hat Y=\hat f(\bx)=\frac{\sum_{i=1}^nY_iK\left(\sum_{j=1}^{p_\text{fun}}\omega_jd_\text{fun}(X_{ij},x_j)+\sum_{j=p_\text{fun}+1}^{p_\text{fun}+p_\text{cat}}\omega_jd_\text{cat}(X_{ij},x_j)\right)}{\sum_{i=1}^nK\left(\sum_{j=1}^{p_\text{fun}}\omega_jd_\text{fun}(X_{ij},x_j)+\sum_{j=p_\text{fun}+1}^{p_\text{fun}+p_\text{cat}}\omega_jd_\text{cat}(X_{ij},x_j)\right)},\]
where $p_\text{fun}$ is the total number of functional covariates, $p_\text{cat}$ the total number of categorical covariates and $p=p_\text{fun}+p_\text{cat}$. 
The weights are estimated by minimizing
$Q(\omega_1,\ldots,\omega_p) = \sum_{i=1}^n (Y_i - \hat{Y}_{(-i)})^2$,
with $\hat Y_{(-i)}$ being the leave-one-out estimate as described above, 
\[\hat Y_{(-i)}=\frac{\sum_{s\neq i}Y_sK\left(\sum_{j=1}^{p_\text{fun}}\omega_jd_\text{fun}(X_{sj},X_{ij})+\sum_{j=p_\text{fun}+1}^{p_\text{fun}+p_\text{cat}}\omega_jd_\text{cat}(X_{sj},X_{ij})\right)}{\sum_{s\neq i}K\left(\sum_{j=1}^{p_\text{fun}}\omega_jd_\text{fun}(X_{sj},X_{ij})+\sum_{j=p_\text{fun}+1}^{p_\text{fun}+p_\text{cat}}\omega_jd_\text{cat}(X_{sj},X_{ij})\right)}\]
in case of MixR. For the minimization we make use of the R  function {\it optim} (\cite{R}) with starting value $(\omega_1,\ldots,\omega_p)=(1,\ldots,1)$, since in this context a brute force optimization routine suffices.

%The minimizing weights for our six different scenarios and $n=100, 500, 1000$ are shown in Figure \ref{fig:Reweights}. To increase comparability between the different models we display normed weights $\frac {\hat\omega_j}{\sum_{k=1}^p\hat\omega_k}$. This can also be interpreted as a separation of the weight estimation (normed weights) and the bandwidth optimization ($h_{\text{opt}}=\frac{h_n^{\text{fun/cat}}}{\sum_{k=1}^p\hat\omega_k}$).
%It can be seen that the selection of relevant predictors works well, as the covariates with influence on $y$ get distinctly higher weights than those without in all scenarios. The sum over the weights for relevant covariates should be approximately one whereas the weights for irrelevant covariates should be close to zero. Both is visible for the simulated data.

%\subsubsection{Prediction performance in regression}\label{ReSimPred}

\subsubsection{Results}\label{ReRe}

\begin{figure}
\begin{centering}
\subfigure{\includegraphics[width=0.41\textwidth]{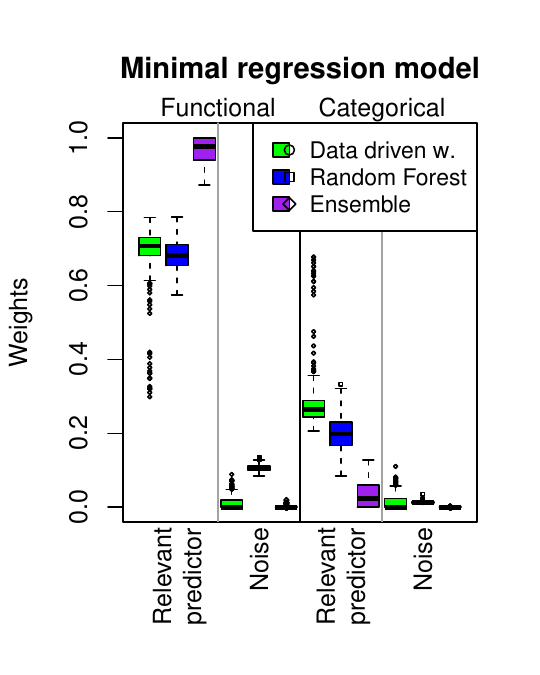}}\hspace{-0.02\textwidth}
\subfigure{\includegraphics[width=0.61\textwidth]{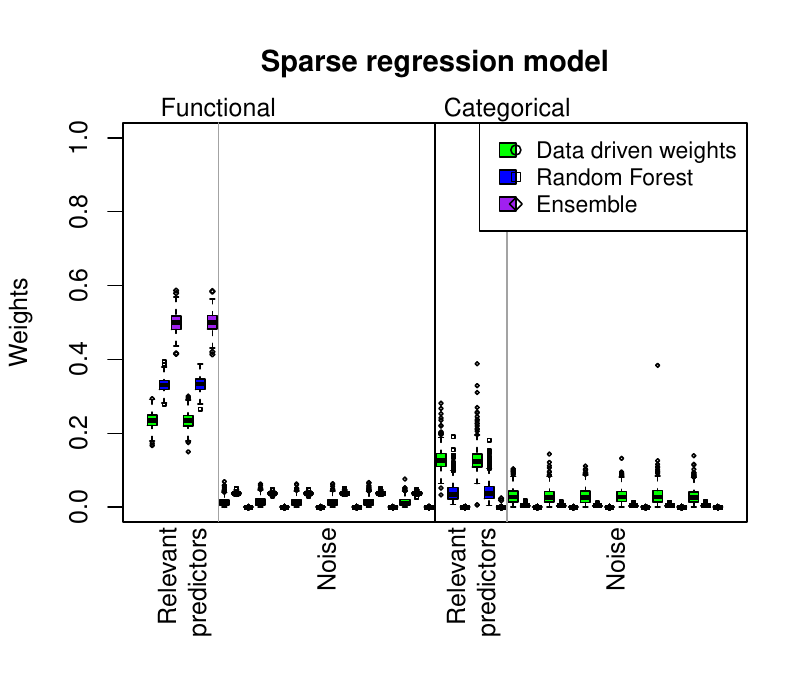}}
\end{centering}
\caption{Normed minimizing weights $\frac{\hat\omega_j}{\sum_{k=1}^p\hat\omega_k}$, variable importance of a random forest, and ensemble weights for model MixR in the minimal (left) and sparse (right) case, respectively. }\label{fig:Reweights}
\end{figure}

The minimizing weights for the minimal as well as the sparse case and sample size $n= 500$ are shown in Figure \ref{fig:Reweights}. They are compared to the relative variable importance of a random forest, as a benchmark apart from kernel-based, nonparametric prediction. After applying a functional principal component analysis (R package {\it refund} by \cite{Rrefund}) on the functional observations we build a random forest using the R function {\it randomForest} (\cite{RrandomForest}). Further we compare our results to the method of \cite{FuchsGertheissTutz2015} (`Ensemble') which was described in Section \ref{introduction}. Although in their paper they only consider categorical responses, their method can also be applied in the regression case. 

To increase comparability between the models we display normed weights $\frac {\hat\omega_j}{\sum_{k=1}^p\hat\omega_k}$. This can also be interpreted as separating the estimation of the weights (normed weights) and optimization of the bandwidth ($h_{\text{opt}}=\frac{h_n^{\text{fun/cat}}}{\sum_{k=1}^p\hat\omega_k}$).
It can be seen that the selection of relevant predictors works well, as the covariates with influence on the response get distinctly higher weights than those without. The sum over the weights for relevant covariates should be approximately one whereas the weights for irrelevant covariates should be close to zero. Both is visible for our procedure. The competing methods get comparable results where the random forest seems to have some difficulties identifying the functional noise and the ensemble approach with detecting the relevant categorical predictors.

\begin{figure}
\begin{centering}

\subfigure{\includegraphics[width=0.5\textwidth]{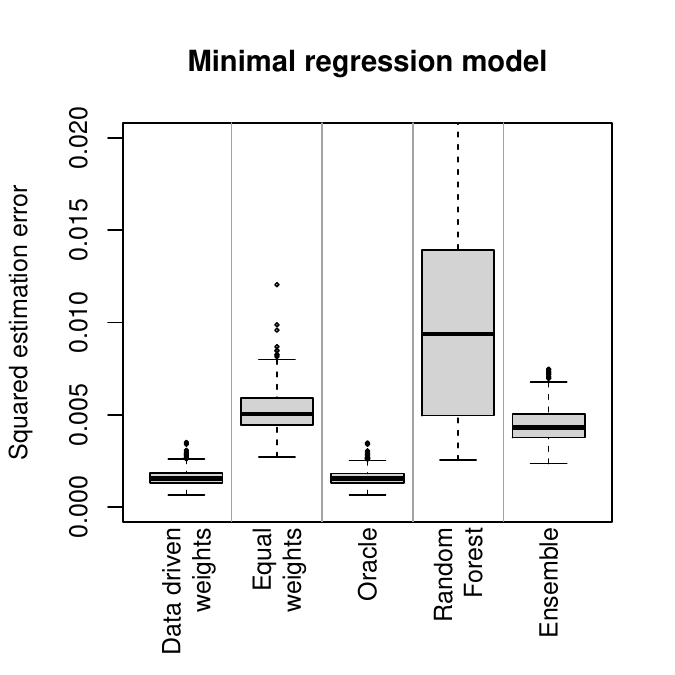}}\hspace{-0.02\textwidth}
\subfigure{\includegraphics[width=0.5\textwidth]{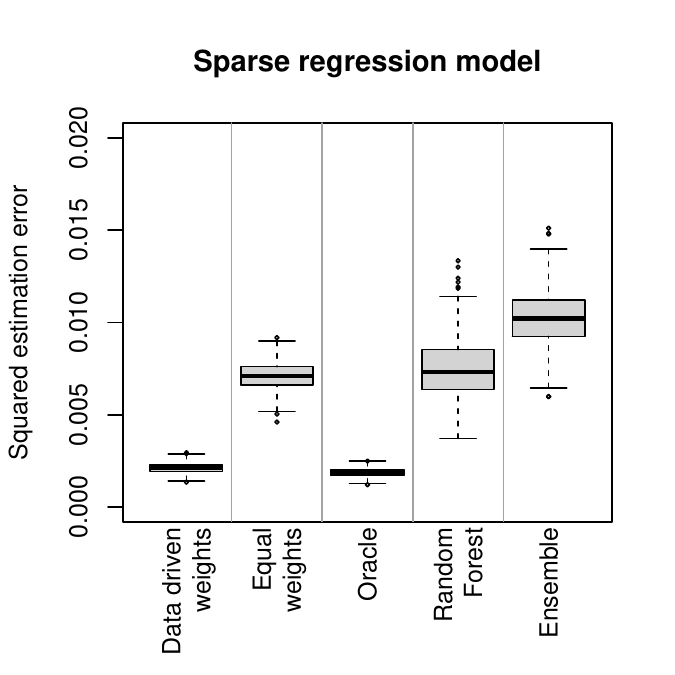}}
\end{centering}
\caption{Estimation performance for model MixR in the minimal (left) and sparse (right) case with no restriction (`data driven weights'), restriction (i, `equal weights') and (ii, `oracle'), and with a random forest and the ensemble approach, respectively. }\label{fig:Remse}
\end{figure}

For further comparison of our prediction results, we also compute the minimizer of $Q$ under the restrictions
\begin{itemize}
\item[(i)] $\omega_1=\omega_2=\ldots=\omega_p$,
\item[(ii)] $\omega_j=0$ for all covariates with no influence on the response. %$j$ with $x_j$ has no influence on $y$.
\end{itemize}
Thus under restriction (ii), which we also call `oracle', we determine the minimizing weights only for the relevant covariates, whereas restriction (i) leads to a single minimizing weight and can be interpreted as determination of a suitable overall/global bandwidth. Note, however, (ii) is only doable in simulations where the truth is known, and no option in practice. In Figure \ref{fig:Remse} the squared estimation error of $\hat f$ is shown, where we display the average over 100 (minimal case) and 10000 (sparse case) $\bx$-values, respectively, and again compare our results to those of a random forest and the method of \cite{FuchsGertheissTutz2015}.  The $\bx$-values are generated randomly in the same way as the covariates. In each of the $500$ replications, new $\bx$-values are generated. The explicit formula to calculate the squared estimation error for each replication is 
$$\frac 1{N}\sum_{l=1}^{N} \left(\frac{\hat f(\bx_l)-f(\bx_l)}{\text{range}(f)}\right)^2,$$
where $N$ is the number of $\bx$-values, $f$ is the true regression function used to generate the data, $\bx_1,\ldots,\bx_{N}$ are the $x$-values (generated at random) and $\text{range}(f)=\max_{l}f(\bx_l)-\min_lf(\bx_l)$.
The results for our procedure are comparable to those under restriction (ii) and better than those under restriction (i), as expected. The competing methods get worse prediction results. Especially compared to the random forest our method is superior.
To get an insight in the influence of the $\bx$-values on the estimation error we ran the simulations also with $\bx$-values that are the same for each replication. The results are almost identical to those with varying $\bx$-values shown in Figure \ref{fig:Remse}.  Only the variance of the estimation errors is slightly larger with varying $\bx$-values (as could be expected).

Another possible way to asses the performance of our procedure would be to look at the (test set) prediction error $Y-\hat f(\bx)=\varepsilon+f(\bx)-\hat f(\bx)$ instead of the estimation error as described above. The results would be similar since the errors $\varepsilon$ are independent of the predictors and thus the mean squared prediction error and the mean squared estimation error only differ in the variance of $\varepsilon$.

\subsection{Classification Problems}\label{ClSim}

\subsubsection{Set-up}
Similar to the regression case, we generate data according to a model (MixC) where we combine functional and categorical predictors. The functional observations are based on those built in model MixR, see Section \ref{ReSim}. Let's call them $X_{ij}^{\text{(Fun)}}$. Then the functional observations for this classification model are $X_{ij}(t)=X_{ij}^{\text{(Fun)}}(t)+0.3\cdot C_{ij}$  with $C_{ij}\sim\mathcal{U}\{0,1\}$. Here $\mathcal{U}$ stands for the discrete uniform distribution.
The categorical covariates are $X_{i(p_{\text{fun}}+1)},$ $\ldots,X_{i(p_{\text{fun}}+p_{\text{cat}})}\sim${\it B}$(0.5)$, such that $p_{\text{fun}}+p_{\text{cat}}=p$. With this,
$$Y_i=(q_{\text{cat}}+1)\cdot(C_{i1}+\ldots+C_{iq_{\text{fun}}})+X_{i(p_{\text{fun}}+1)}+\ldots+X_{i(p_{\text{fun}}+q_{\text{cat}})}+1,$$
$q_{\text{fun}}\leq p_{\text{fun}}$, $q_{\text{cat}}\leq p_{\text{cat}}$, and thus $G=(q_{\text{fun}}+1)\cdot(q_{\text{cat}}+1)$.

As before we compare minimal (*.m) and sparse (*.s) cases, i.\,e.,   $q_{\text{fun}}=q_{\text{cat}}=1$, $p_{\text{fun}}=p_{\text{cat}}=2$ (*.m) and $q_{\text{fun}}=q_{\text{cat}}=2$, $p_{\text{fun}}=p_{\text{cat}}=8$ (*.s). The results are based on 500 replications.
We use again the one-sided Picard kernel as described in Section \ref{ReSim}.
%\begin{figure}
%\begin{centering}
%\subfigure{\includegraphics[width=0.33\textwidth]{FunC01.pdf}}\hspace{-0.02\textwidth}
%\subfigure{\includegraphics[width=0.33\textwidth]{FunC03.pdf}}\hspace{-0.02\textwidth}
%\subfigure{\includegraphics[width=0.33\textwidth]{FunC07.pdf}}
%%\subfigure{\includegraphics[width=0.49\textwidth]{FunC1.pdf}}
%\end{centering}
%\caption{Examples for functional observations from model (FunC) for $c=0.1,0.3,0.7$ (from left to right). The green dashed lines represent $X_{ij}$ with $C_{ij}=2$, the blue solid lines those with $C_{ij}=1$ and the red dotted lines are realizations of $X_{ij}$ with $C_{ij}=0$.  }\label{fig:FunC}
%\end{figure}
In contrast to the regression case, however, we use a pre-estimator for the weights instead of a starting value for the bandwidth. Thus, we set
\begin{eqnarray*}
d_{\text{fun}}(X_{ij},x_j)&:=&\frac 1{c_j^\text{fun}}\sqrt{\int(X_{ij}(t)-x_j(t))^2dt},\\
d_{\text{cat}}(X_{ij},x_j)&:=&\frac 1{c_j^\text{cat}}I\{X_{ij}\neq x_j\},
\end{eqnarray*}
with norming constants $c_j^\text{fun}=\sqrt{\int \frac1{n-1}\sum_{l=1}^n(X_{lj}(t)-\frac 1n\sum_{k=1}^nX_{kj}(t))^2dt}$, \\$c_j^\text{cat}=\sqrt{\frac1{n-1}\sum_{l=1}^n(X_{lj}-\frac 1n\sum_{k=1}^nX_{kj})^2}$, respectively, 
and determine the starting values $(\hat\omega_1^{\text{pre}},\ldots,\hat\omega_p^{\text{pre}})$ for minimizing
$Q(\omega_1,\ldots,\omega_p) = \sum_{i=1}^n\sum_{g=1}^G (I\{Y_i=g\} - \hat{P}_{g(-i)})^2$  by
\[\hat\omega_j^{\text{pre}}:=\arg\min_{\omega}\sum_{i=1}^n\sum_{g=1}^G (I\{Y_i=g\} - \hat P_{g(-i)}^{\text{pre}}(j,\omega))^2\]
with
\[\hat P_{g(-i)}^{\text{pre}}(j,\omega)=\frac{\sum_{s\neq i}I\{Y_s=g\}K\left(\omega d_\text{fun/cat}(X_{sj},X_{ij})\right)}{\sum_{s\neq i}K\left(\omega d_\text{fun/cat}(X_{sj},X_{ij})\right)}\]
where $d_\text{fun/cat}$ means that $d_\text{fun}$ or $d_\text{cat}$ is used according to the type of the $j$th predictor. Of course it would have been possible to use the pre-estimator in the regression framework as well. Since in the regression case, however, there are well-known rules of thumb at hand for the bandwidth / weights selection, it may be preferable to use those to reduce computation time.

%In Figure \ref{fig:Clweights} the minimizing normed weights for model (CatC), (FunC) (with $c=0.3$) and (MixC) for $n=100,500,1000$ are displayed. The performance regarding the variable selection is very satisfying. In Figure \ref{fig:ClweightsFun} the weight estimation results for model (FunC) for other values of $c\in(0,1]$ are shown in the minimal case. As expected the variable selection works better for larger $c$. 

%\begin{figure}
%\begin{centering}
%\subfigure{\includegraphics[width=0.49\textwidth]{WfunMinCl01.pdf}}
%\subfigure{\includegraphics[width=0.49\textwidth]{WfunMinCl07.pdf}}
%\end{centering}
%\caption{Normed minimizing weights $\frac{\hat\omega_j}{\sum_{k=1}^p\hat\omega_k}$ for model (FunC)  in the minimal case for $c=0.1$ (left-hand side) and $c=0.7$ (right-hand side). }\label{fig:ClweightsFun}
%\end{figure}

%\subsubsection{Prediction performance in classification}

\subsubsection{Results}\label{ClRe}

\begin{figure}
\begin{centering}
\subfigure{\includegraphics[width=0.41\textwidth]{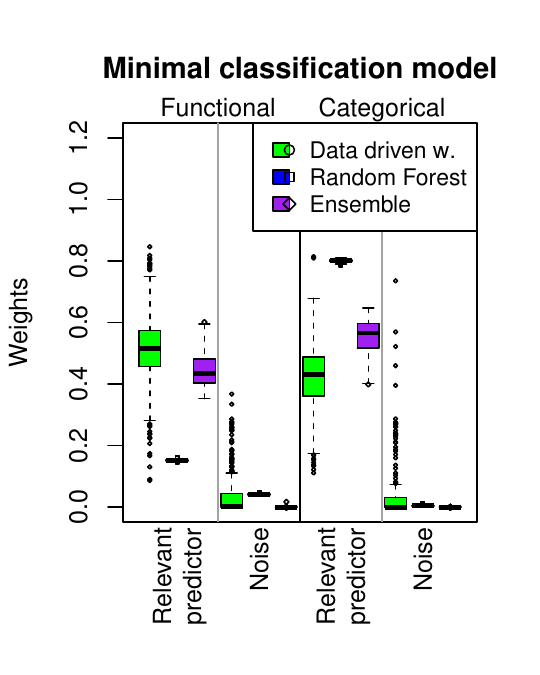}}\hspace{-0.02\textwidth}
\subfigure{\includegraphics[width=0.61\textwidth]{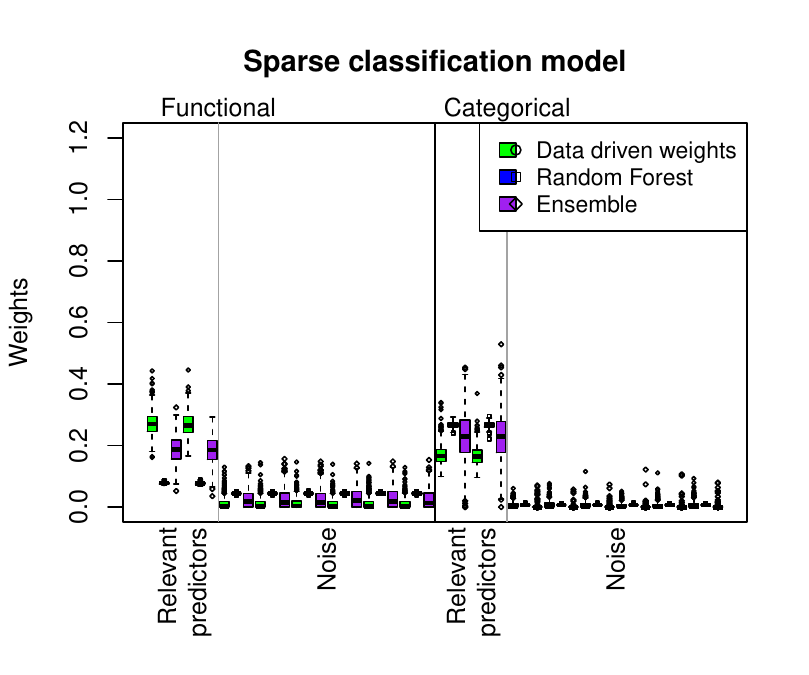}}
\end{centering}
\caption{Normed minimizing weights $\frac{\hat\omega_j}{\sum_{k=1}^p\hat\omega_k}$, variable importance of a random forest, and ensemble weights for model MixC in the minimal (left) and sparse (right) case, respectively.. }\label{fig:Clweights}
\end{figure}

As described in Section \ref{ReRe} we again compare our results to those of a random forest and the ensemble method of \cite{FuchsGertheissTutz2015}. In Figure~\ref{fig:Clweights} the minimizing normed weights for model MixC and $n=500$ are displayed. The performance regarding the variable selection is very encouraging and for the functional covariates clearly better than that of the random forest. %In Figure \ref{fig:ClweightsFun} the weight estimation results for model (FunC) for other values of $c\in(0,1]$ are shown in the minimal case. As expected the variable selection works better for larger $c$. 

\begin{figure}
\begin{centering}
\subfigure{\includegraphics[width=0.5\textwidth]{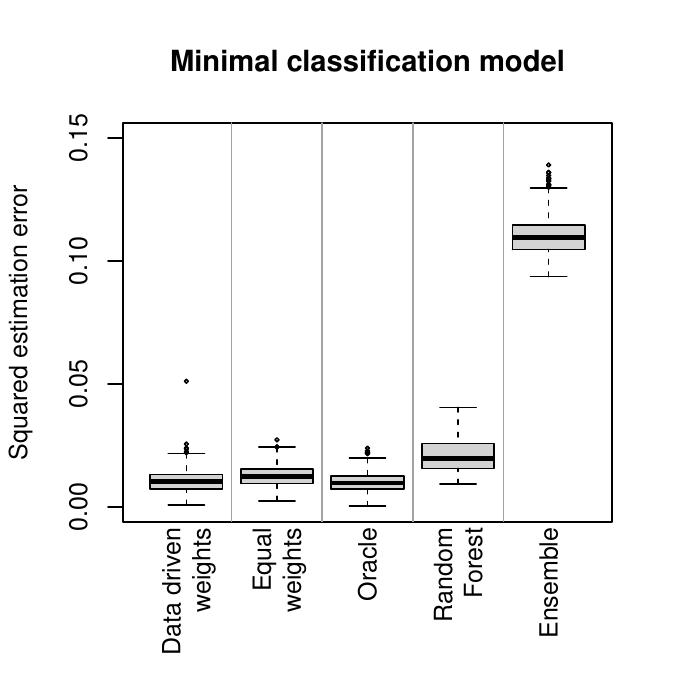}}\hspace{-0.02\textwidth}
\subfigure{\includegraphics[width=0.5\textwidth]{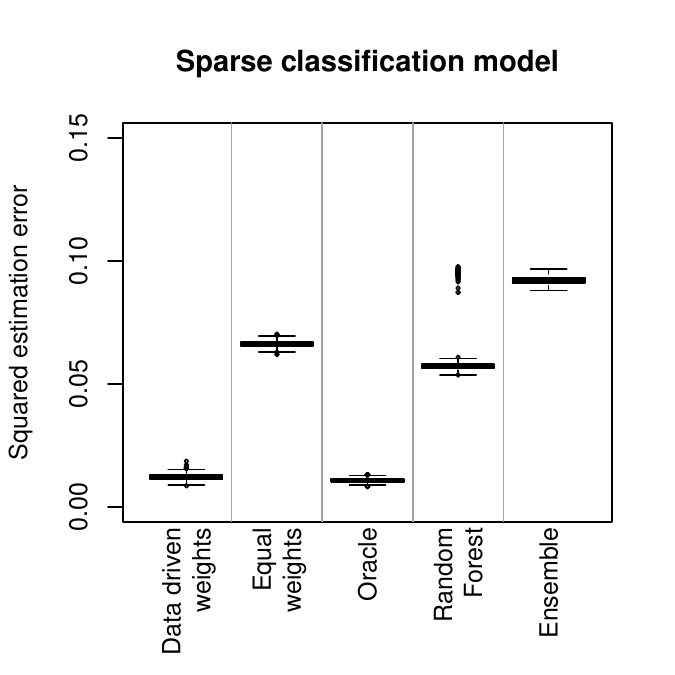}}
\end{centering}
\caption{Estimation performance for model MixC in the minimal (left) and sparse (right) case with no restriction (`data driven weights'), restriction (i, `equal weights') and (ii, `oracle'), and with a random forest and the ensemble approach, respectively. }\label{fig:Clmse}
\end{figure}

The estimation performance of our procedure is shown in Figure \ref{fig:Clmse}, where we display the squared error of $\hat P_g$ and compare it to the results under restriction (i) and (ii) as described in Section \ref{ReRe}. Furthermore our approach is compared with the competing methods `Random Forest' and `Ensemble'. For new $\bx$-values (that are generated in the same way as the observations from the training set), we predict the posterior probability with the random forest, the ensemble and with our $\hat P_g$ with the estimated weights, respectively. The data for the boxplots is calculated on test sets with $N=100$ (minimal case) and $N=1000$ (sparse case) as the Brier Score
\[\frac 1N\sum_{l=1}^N\frac 1G\sum_{g=1}^G(\hat P_g(\bx_l)-I\{y(\bx_l)=g\})^2,\]
where $y(\bx)$ is the response (class) resulting from the predictor $\bx$.
 $y(\bx)$ are built in the same way as for the training observations.
Similar to the regression case, the results achieved with new $\bx$-values for each replication and those with the same $\bx$-values in all replications are comparable. We display the results with varying $\bx$-values. 
It can be seen that the prediction works well and clearly better than the random forest and the ensemble method. Further in the sparse case, the results with data driven weights are much better than those with equal weights, which confirms the good variable selection/weighting performance.

As additional information we display the missclassification rate as an arithmetic mean over
\[\frac 1N\sum_{l=1}^NI\left\{\arg\max_{g\in\{1,\ldots,G\}}\hat P_g(\bx_l)\neq y(\bx_l)\right\}.\]
The results are summed up in Table~\ref{tab:missCl}. They confirm the good performance shown in Figure \ref{fig:Clmse}, especially that our procedure works much better than the random forest and the ensemble approach. In particular, the superior performance of using weights within the kernel as proposed here instead of combining individual, covariate-specific nearest neighbor predictions as an ensemble can be explained as follows. Whereas the nonparametric, kernel-based approach as presented in Section~\ref{Methods} is able to handle/incorporate interactions between predictors, this is hardly possible by simply combining predictions each based on a single covariate only by means of a weighted average (as done with the ensemble). Furthermore, nearest neighbor predictions that use a single, binary predictor only, tend to be poor (which also affects the ensemble at least to some degree). As a result, the nearest neighbor ensemble approach may not be the way to go with categorical predictors that only have a small number of categories. The results for further models we simulated can be found in the online supplement.

\begin{table}
\begin{centering}
\begin{tabular}{c||c|c|c|c|c}
Model  & Data driven w. & Equal weights & Oracle & Random forest & Ensemble\\
\hline
\hline
(MixC.m)
 & {\color{teal}0.03 (0.02)} & {\color{teal}0.03 (0.03)} & {\color{teal}0.03 (0.02)} & 0.22 (0.38) & 0.25 (0.09)\\
\hline
(MixC.s) 
 & {\color{violet}0.07 (0.01)} & 0.44 (0.02) & {\color{teal}0.06 (0.01)} & 0.21 (0.24) &  0.72 (0.04)
\end{tabular}
\caption{Missclassification rates as arithmetic mean (and standard deviation) with no restriction (`Data driven weights'), restriction (i) (`Equal weights'), restriction (ii) (`Oracle'), and with a random forest and the ensemble method, respectively. The values in teal are the lowest and the values in violet the second to lowest in each row. }\label{tab:missCl}
\end{centering}
\end{table}

\section{Application to Real World Data}\label{RealData}

Finally, we apply our procedure to some real world data. The first one is the \emph{ArabicDigits} data described in the Introduction. Further we consider trajectory data from a psychological experiment with virtual reality devices, as well as another three benchmark data sets. The first one is data from a medical survey investigating the response of patients to drug therapy. 
The second one is data from a psychological survey investigating the effect of different movies on the motivational state of participants. The third one is a well-known data set on the housing situation in Copenhagen.

\subsection{Speech Recognition}

As an example for a multi-class classification problem with multiple functional predictors we consider the data set {\it ArabicDigits} from the R-package {\it mfds} by \cite{Rmfds}, see Section \ref{introduction}.
%This dataset contains time series of 13 Mel Frequency Cepstrum Coefficients (MFCCs) corresponding to spoken Arabic digits. MFCCs are very common for speech recognition, see \cite{KoolagudiRastogiRao2012} for a detailed explanation.
Each time series in the 13 speech features contains 93 data points and the number of  time series is 8800 (10 digits x 10 repetitions x 88 speakers) in total. We split the data in each group randomly in a training and a test set in the relation 70/30. Thus we estimate our weights based on $n=6160$ observations with $p=13$, $G=10$  and $T=93$.

\begin{figure}
\begin{centering}
\includegraphics[width=\textwidth]{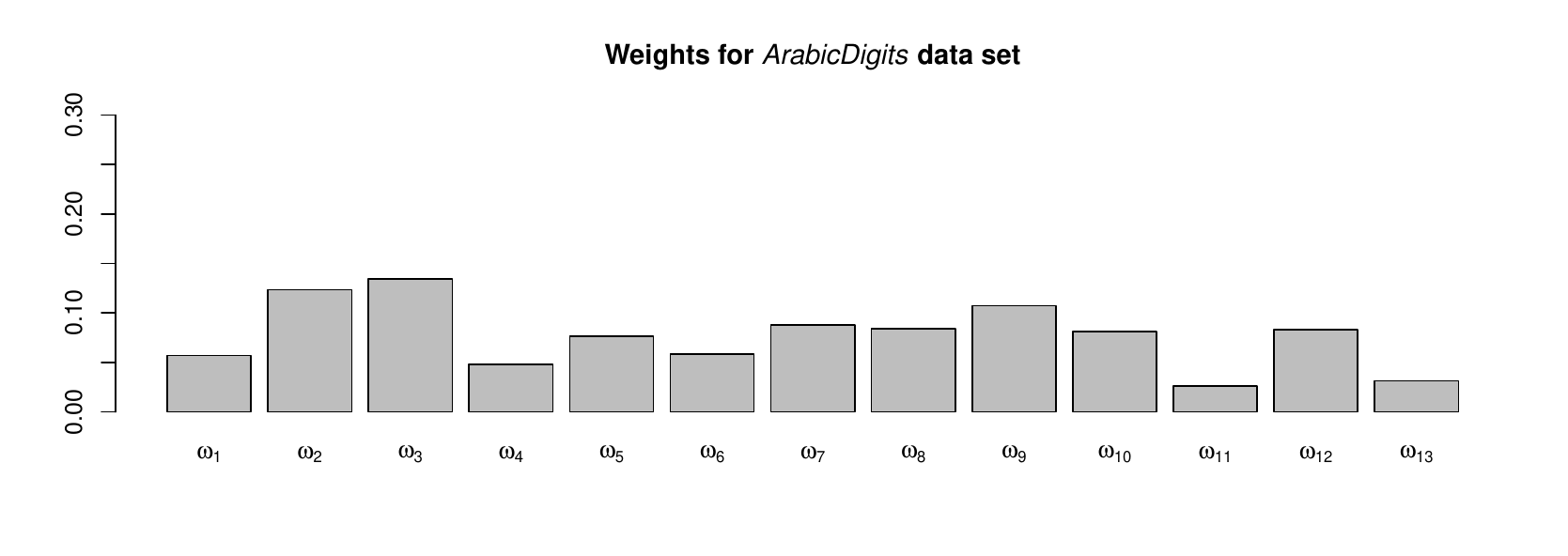}
\caption{Estimated weights (normed) for the 13 MFCCs in the {\it ArabicDigits} data set.}\label{fig:ADweights}
\end{centering}
\end{figure}

The results show that all 13 MFCCs are relevant as expected. The 13 normed weights are all of the same size around $1/13$, see Figure \ref{fig:ADweights}.
Further the prediction results for the test data set (2640 observations) are almost perfect as can be seen in Table \ref{tab:ADres}. This very good prediction performance is comparable to results of other procedures applied on this data set. For instance, \cite{GoreckiLuczak2015} model the data as multivariate time series and use a 1NN classification where the distance measure is based on dynamic time warping. A (parametric) functional multivariate regression approach for multi-label classification is used by \cite{KrzyskoSmaga2017}. 
In \cite{MoellerGertheiss2018} a classification tree is applied. The authors choose arbitrarily two out of the 10 digits to make the problem a binary classification task. They all get very good prediction results for this data set as well. 

\begin{table}
\begin{centering}
\begin{tabular}{c|cccccccccc}
& 0 & 1 & 2 & 3 & 4 & 5 & 6 & 7 & 8 & 9 \\
 \hline
 0 & 261 &0 &0 &0 &0 &2 &1 &0 &0 &0\\
 1& 0& 264& 0& 0& 0& 0& 0& 0& 0& 0\\
 2 &  0&   0 &263&   0&   0&   0&   0&   0&   1&   0\\
  3   &0 &  0  & 0 &264 &  0  & 0 &  0&   0&   0&   0\\
  4   &0  & 0   &0  & 0 &264&   0&   0   &0   &0   &0\\
  5   &0   &0   &0   &1 &  0 &263  & 0 &  0&   0&   0\\
  6   &2   &0   &0   &0   &0   &0 &262&   0   &0   &0\\
  7   &0   &0   &0   &0   &0   &0 &  0 &261   &0&   3\\
  8   &0   &0   &0   &0   &0   &0   &0   &0 &264   &0\\
  9   &0   &0   &0   &0   &0   &0   &0 &  2   &0& 262
\end{tabular}
\caption{Classification results for the {\it ArabicDigits} data set as a contingency table of the true (rows) and the estimated (columns) classes. }\label{tab:ADres}
\end{centering}
\end{table}

\subsection{Virtual Reality Movement Data}

\begin{figure}
\begin{centering}
%\subfigure{\includegraphics[width=0.49\textwidth]{TraLHgli9.pdf}}
\subfigure{\includegraphics[width=0.49\textwidth]{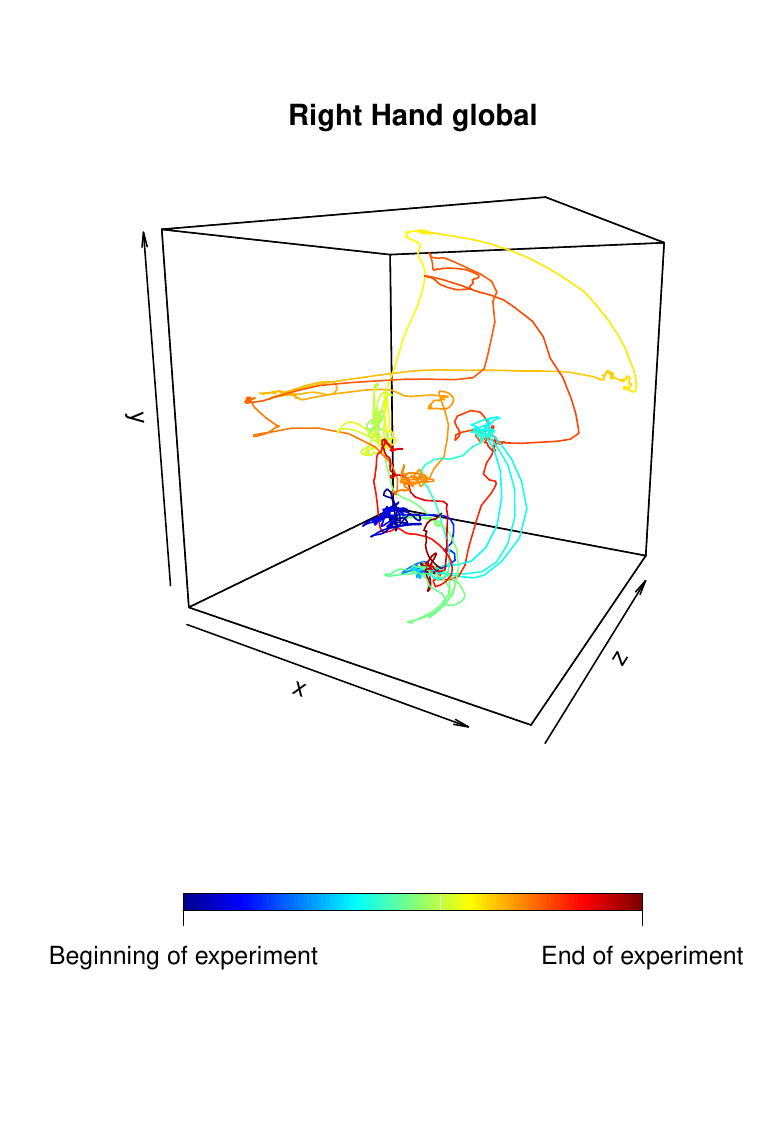}}
\subfigure{\includegraphics[width=0.49\textwidth]{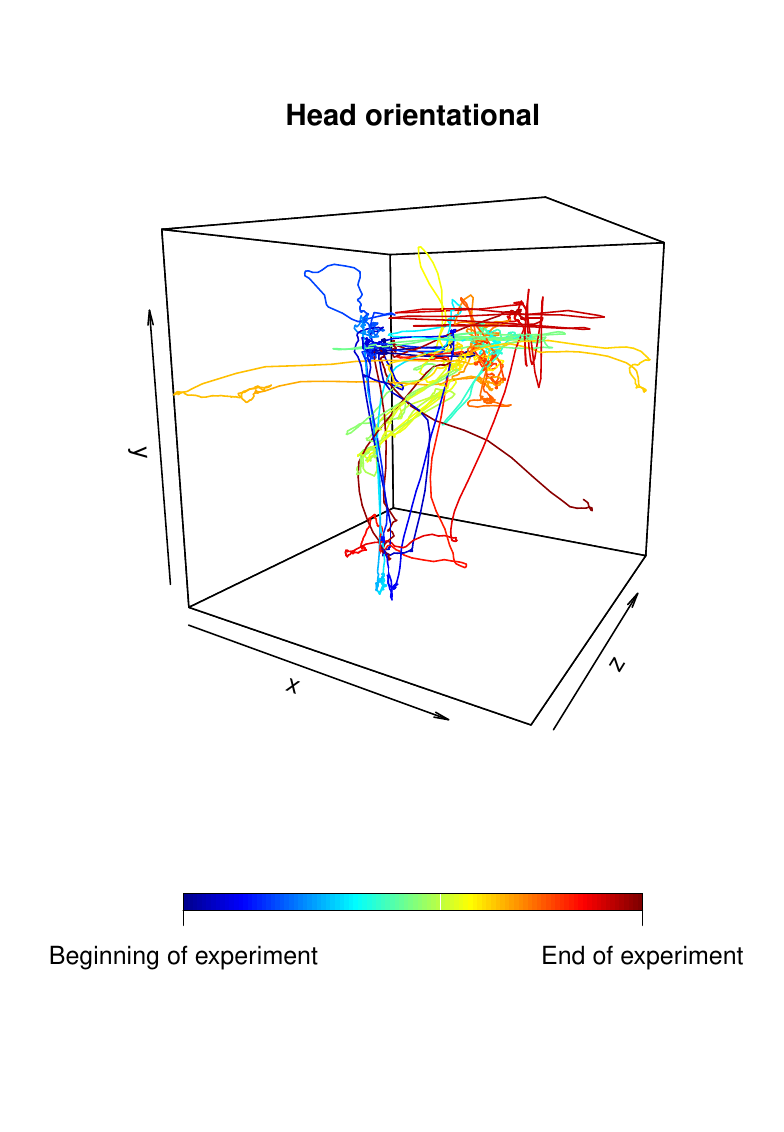}}
\end{centering}
\caption{Examples for trajectories from one female individual. The orientational movement of the head (right chart) can be interpreted as follows: An increase in the x-component refers to looking to the right; a decrease to looking to the left. An increase in the y-component refers to looking up; a decrease to looking down. An increase in the z-component refers to tilting the head to the right; a decrease to tilting it to the left.}\label{fig:trajec}
\end{figure}
Besides `classical', one-dimensional functional data, also other types of functional data such as 3-dimensional trajectories are getting more and more attention; see, e.g., \cite{Fernandez-FonteloEtal2021}. The data set considered in the paper at hand contains 3-dimensional movement data of the hands and head of participants in a psychological experiment, compare \cite{VogelEtal2022}. The participants were asked to perform guided upper body exercises like stretching their arms or embracing themselves. Furthermore, the participants were given a virtual reality headset and two joysticks, one for each hand. With these devices the movements of the hands and the head were recorded. The movements are recorded as `global', `local' and `orientational', where `global' describes the position in the lab, `local' the position relative to the position of the feet and `orientational' records rotational motions; see \cite{VogelEtal2022} for a more detailed description of the experimental setting,  and \cite{VahleTomasik2021} for a similar experiment, with the focus being on memory performance, physical strength and endurance. Figure \ref{fig:trajec} shows some example trajectories. It is easy to identify from the left chart the time point when the participant is asked to raise her hands and to build the letter `T' directly afterwards. The virtual reality that was created for the participants, was an avatar to mimic the movements. The participants were all young, while the avatars were either young or elderly people. One of the questions of this psychological experiment was whether the experimental condition, that is, the class of the avatar (young vs.~elderly person) could be reconstructed from the movement data. Thus we have a binary classification problem with multiple 3-dimensional functional predictors (trajectories).

The task is to predict/identify the class of the avatar (young vs.~elderly person) using the movement data. To allow for the multi-dimensional functional predictors (the trajectories), we set 
\[d(X_{ij},x_j)=\frac{1}{c_j}\sqrt{\sum_{r=1}^3\int(X_{ij}^{(r)}(t)-x_j^{(r)}(t))^2dt}\]
where $X^{(r)}(t)$ describes the $r$-th component of $X(t)$ and 
\[c_j=\sqrt{\sum_{r=1}^3\int \frac1{n-1}\sum_{l=1}^n(X_{lj}^{(r)}(t)-\frac 1n\sum_{k=1}^nX_{kj}^{(r)}(t))^2dt}.\]

%We use the same data that is examined in \cite{VogelEtal2021} where several supervised learning procedures are applied to assign the movement patterns of both hands to the underlying avatar class. In contrast to our trajectory approach they treat each dimension of the movement observations separately as a one-dimensional functional observation. A similar experiment is considered in \cite{VahleTomasik2021}, with the focus being on memory performance, physical strength and endurance.

\begin{figure}
\begin{centering}
\includegraphics[width=0.8\textwidth]{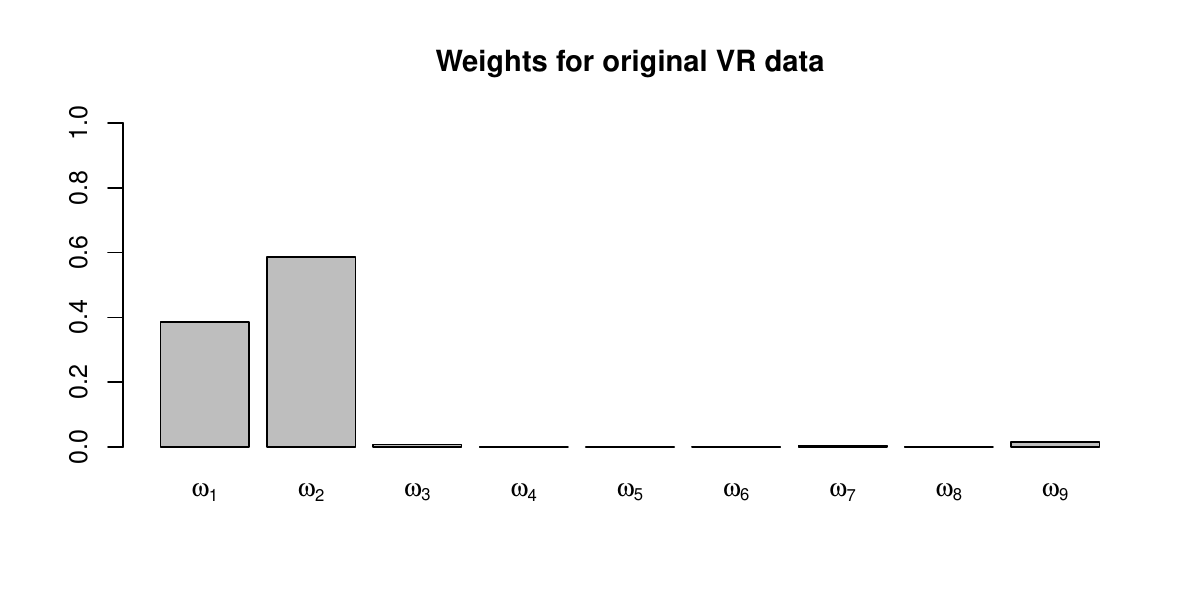}
%\subfigure{\includegraphics[width=0.39\textwidth,height=0.3\textwidth]{WVR6.pdf}}
\end{centering}
%\caption{Estimated weights (normed) for the 9 (6) predictors (3 (2) body parts x 3 coordinate systems) in the VR data set with the original (left-hand side) and the trimmed (right-hand side) data. On the left-hand side, the first 3 weights belong to the head movements (global, local, orientational), the middle 3 weights to the left hand movements, and the last 3 weights to the right hand movements. On the right-hand side, the first/last 3 weights belong to the left and right hand movements (global, local, orientational), respectively.}\label{fig:VRweights}
\caption{Estimated weights (normed) for the 9 predictors (3  body parts x 3 coordinate systems) in the VR data set. The first 3 weights belong to the head movements (global, local, orientational), the middle 3 weights to the left hand movements, and the last 3 weights to the right hand movements.}\label{fig:VRweights}
\end{figure}

%In the experiment we consider the participants are asked to perform guided upper body exercises like stretching their arms or embracing themselves.
%The movements are recorded as global, local and orientational where global describes the position in the lab, local the position relative to the position of the feet and orientational records rotational motions, see \cite{VogelEtal2021} for a more detailed description of the experimental setting. In Figure \ref{fig:trajec} an example for the trajectories that are employed as predictors is shown. It is easy to identify from the chart the time point when the participant is asked to raise his hands and to build the letter T directly afterwards.
In total there is data from $n=72$ participants available and the movements are tracked with a frequency of 10 Hz, resulting in patterns consisting of $T=4970$ time points per coordinate. The data is available at \texttt{\small https://osf.io/h53rk/}. We estimate the weights with all available observations. The results are shown in Figure \ref{fig:VRweights}. %After the estimation of the weights we predict the class for each observation in a leave-one-out manner with the weights determined before. The prediction results are displayed in Figure \ref{fig:VRres}, left-hand side. 
It can be seen that the first two predictors (global and local position of the head) are weighted distinctly higher than the following 7 predictors. %This corresponds to the prediction results that are clearly better with the data driven weights than with equal ones. 
The reason for this effect, however, became clear after some closer inspection of the data as there is an artificial additive shift between groups for the local head data. Due to a coding error, the reference point for the local head data is different between groups. This shift is not apparent in the global head data and thus, since both components describe the same movements, their combination is a good predictor. Although this effect is only an artifact, we nevertheless present the results for the entire data set since they confirm the good performance of our procedure in terms of variable selection.

\subsection{Impact of gene expressions on the responses to drug therapy}

This real data example is considered due to its potential for variable selection. The data set contains 
gene expressions of $p=76$ genes which are mainly related to the immune system from $n=53$ multiple sclerosis patients that were treated with interferon beta (IFN-$\beta$). After an observation period of 2 years the patients were categorized into good and poor responders. Thus we deal with a binary classification problem with multiple functional predictors. 
The gene expression levels were measured at the beginning of the treatment and after 3, 6, 9, 12, 18, and 24 months. Since there are missing values the number of time points range from $T=4$ to $T=7$.
In \cite{BaranziniEtal2004} this data set is explained and examined elaborately including a longitudinal analysis of the genes responder effect using a repeated-measures analysis of variance.
\cite{KayanoEtal2016} take this data set as an application example for their method of differential analysis for time course gene expression profiles. They apply a functional logistic model to identify the genes with dynamic alterations in good/poor responders. The same data has also been analyzed by \cite{HiroseEtal2007} who applied clustering algorithms.

\begin{figure}
\begin{centering}
\includegraphics[width=0.8\textwidth]{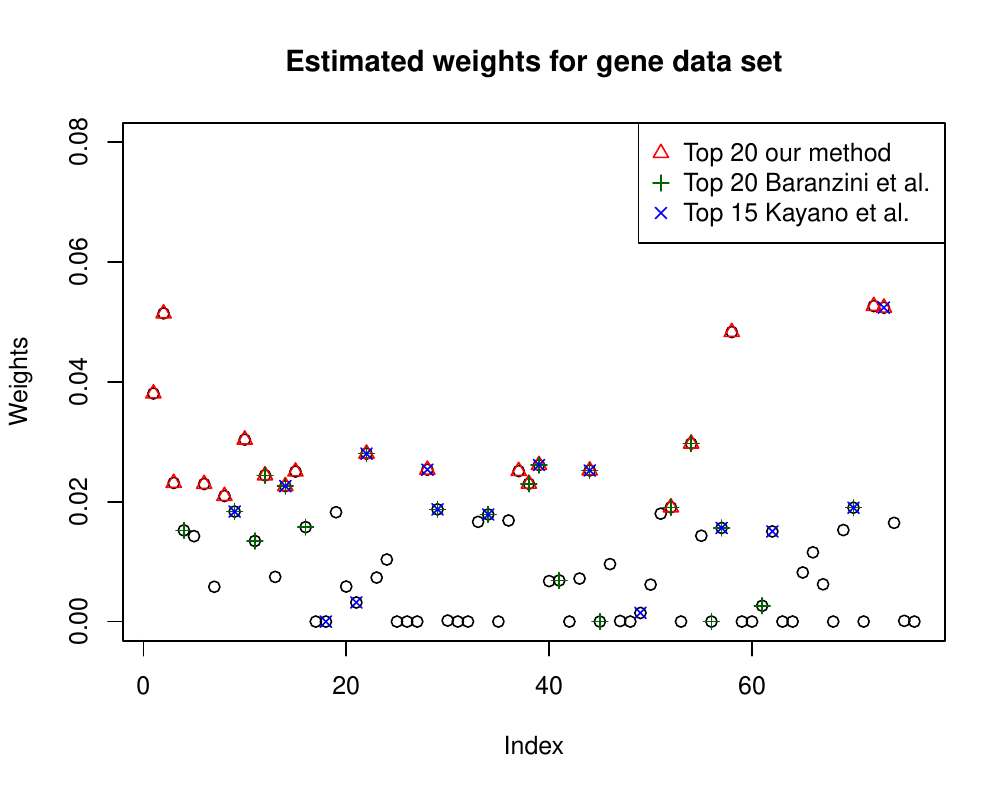}
\caption{Estimated weights (normed) for all 76 weights of the gene data set. The red triangles indicate the 20 genes weighted highest by our method. The darkgreen $+$ mark the top 20 genes in \cite{BaranziniEtal2004}. The blue $\times$ stand for the top 15 genes in \cite{KayanoEtal2016}.  }\label{fig:Bioweights}
\end{centering}
\end{figure}

We estimate the weights with all available observations ($n=53$) and afterwards predict the class for each observation in a leave-one-out manner with weights that are newly estimated with all but the one observation that shall be predicted. In Figure \ref{fig:Bioweights} the estimated weights are shown and compared to the most significant genes for predicting the responder effect determined by \cite{BaranziniEtal2004} and \cite{KayanoEtal2016} respectively. In addition, in Table \ref{tab:Bioweights} the 20 genes weighted highest by our method are listed. It can be seen that 8 (resp. 6)  of the 20 genes are also part of the 20 (resp. 15) most significant ones of \cite{BaranziniEtal2004} (resp.~\cite{KayanoEtal2016}). \cite{BaranziniEtal2004} and \cite{KayanoEtal2016} match in 9 genes.
%The prediction results are displayed in Figure \ref{fig:Biores}. It can be seen that the results with the data driven weights are better than those with equal weights which confirms that not all predictors are relevant.

%alle drei zusammen: 4
%rot und gr�n: 4
%rot und blau: 2
%blau und gr�n: 5
%rot alleine: 10
%gr�n alleine: 7
%blau alleine: 4

\begin{table}
\begin{centering}
\begin{tabular}{c|c|c}
Gene &\  Normed weight\  & Selected by other methods\\
\hline\hline
CD22 & 0.0526\\
CD69 & 0.0524 & $\times$\\
IFNaR2 & 0.0514\\
ITGA & 0.0483\\
IFNaR1 & 0.0381\\
IL12Rb1 & 0.0304\\
IRF4 & 0.0297& $+$\\
GRB2 & 0.0280 & $+$ $\times$\\
CASPASE5 & 0.0261 & $+$ $\times$\\
STAT4 & 0.0253 & $\times$\\
CASPASE10 & 0.0252 & $+$ $\times$\\
CASPASE3 & 0.0251\\
NFATC2b & 0.0250\\
TYK2 & 0.0244 & $+$\\
IL10 & 0.0231\\
CASPASE4 & 0.0230 & $+$\\
 IL10Rb & 0.0230\\
JAK2 & 0.0226 & $+$ $\times$ \\
IFNgRa & 0.0210\\
IRF2 & 0.0191& $+$
\end{tabular}
\caption{Genes weighted highest by our method (top 20) with a comparison to other methods, where $+$ means the gene is also part of the top 20 in \cite{BaranziniEtal2004} and $\times$ stands for part of the top 15 in \cite{KayanoEtal2016}. }\label{tab:Bioweights}
\end{centering}
\end{table}

%\begin{figure}
%\begin{centering}
%\includegraphics[width=0.8\textwidth]{MSEBioHist.pdf}
%\caption{Histogramms of the squared errors of the estimated posterior probabilities for estimators with data driven weights (left-hand side) and with equal weights (right-hand side), respectively, for the gene data set. }\label{fig:Biores}
%\end{centering}
%\end{figure}

\subsection{Effect of Movies on Motivational State }
 
This data set is considered as an example for multi-class classification with categorical predictors. 
It is called {\it msq} and is included in the R-package {\it psychTools} by \cite{RpsychTools}. MSQ stands for motivational state questionnaire in which participants were asked to indicate their current standing on a four-point scale from $0$ (`Not at all') to $3$ (`Very much') for 72 emotions like `afraid', `angry', `cheerful', `happy', `relaxed', etc. The whole data set contains data from 38 studies with different focuses. We were interested in the effect of different movies shown to the participants and thus used the data from the studies `FLAT' and `Maps'. The movies shown were 9 minute clips from 1) a BBC documentary on British troops arriving at the Bergen-Belsen concentration camp, 2) a scene from the horror movie `Halloween', 3) a documentary about lions on the Serengeti plain, and 4) a scene from the comedy `Parenthood'. Our aim was then to predict the movie a participant has seen based on his or her MSQ before and after seeing the clip. For this sake we built the differences between the ratings on the MSQ that was filled out after seeing the movie and the ratings on the MSQ from before the movie, and used these values as categorical predictors. Thus a value of e.\,g.\ $-3$ means that this emotion was rated as $3$ (`Very much') before the movie and as $0$ (`Not at all') after the movie.

\begin{figure}
\begin{centering}
\includegraphics[width=\textwidth]{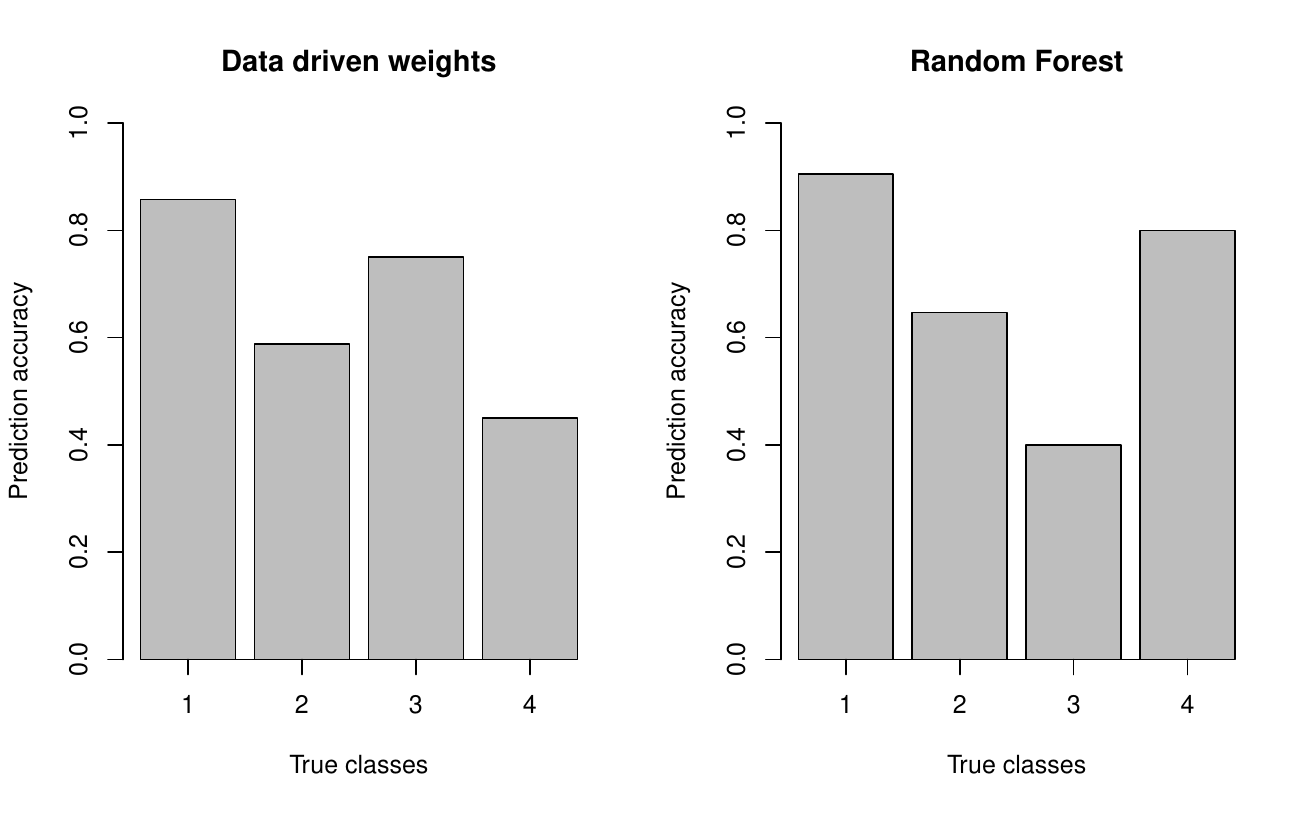}
\caption{Prediction accuracy per class for the {\it MSQ} data set with our procedure (data driven weights) and with a random forest. }\label{fig:MSQRes}
\end{centering}
\end{figure}

In total our weight estimation and prediction was based on $n=188$ training observations with $G=4$ and $p=72$, where each of the 72 predictors is a categorical variable with values in $\{-3,-2,-1,0,1,2,3\}$. Figure \ref{fig:MSQRes} shows the satisfying prediction results based on a 70/30 split in training and test data for our procedure and for a random forest in comparison.

\begin{figure}
\begin{centering}
\includegraphics[width=0.8\textwidth]{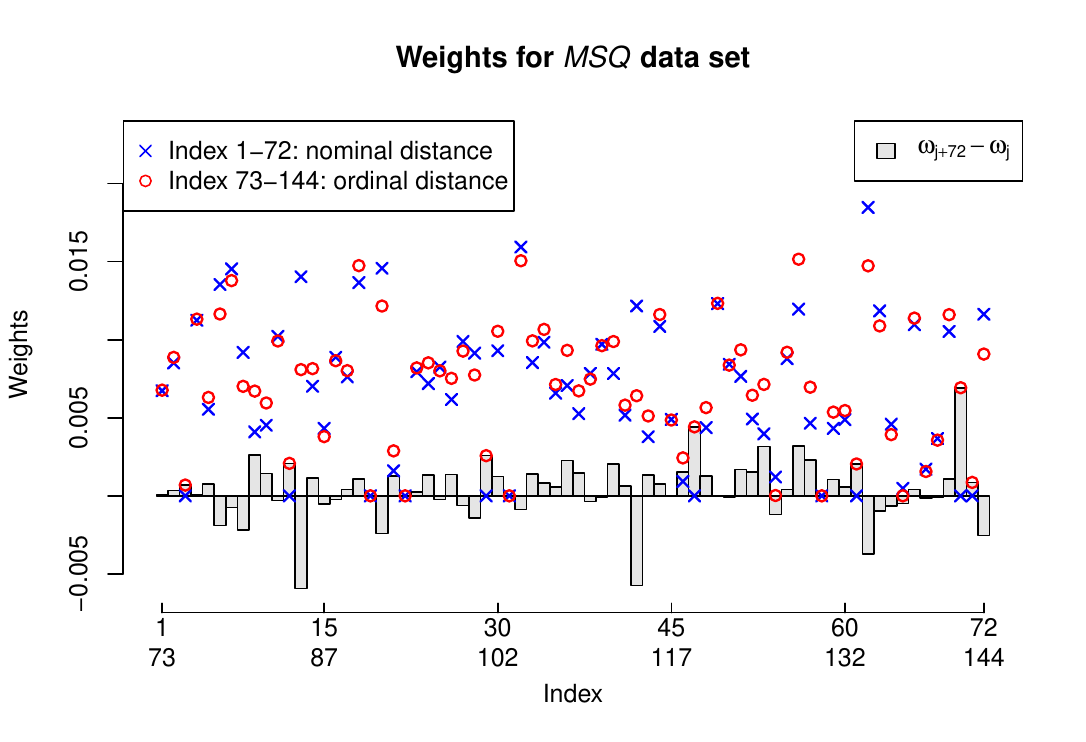}
\caption{Estimated weights for a combination of two copies of the {\it MSQ} data set with different distance measures, namely $d_{\text{nom}}$ (`nominal distance') and $d_{\text{ord}}$ (`ordinal distance'). The grey bars indicate the differences between the weights that correspond to the ordinal and those that correspond to the nominal distance measure.}\label{fig:MSQWeights}
\end{centering}
\end{figure}

Since the predictor-data is ordinal we also considered the distance measure 
\[d_{\text{ord}}(X_{ij},x_j)=\frac 1{c_j^{\text{ord}}}|X_{ij}-x_j|\]
in addition to 
\[ d_{\text{nom}}(X_{ij},x_j) =\frac 1{c_j^{\text{nom}}}\cdot \left\{ \begin{array}{ll}
0 & \mbox{ if } X_{ij} = x_j\\
1 & \mbox{ if } X_{ij}\neq x_j\end{array}
\right.\]
which is similar to the distance measure introduced in \eqref{distcat}. The norming constants are data dependent: 
\[c_j^{\text{ord}}=\sqrt{ \frac1{n(n-1)-1}\sum_{s=1}^{n}\sum_{t\neq s}\left(|X_{sj}-X_{tj}|-\frac 1{n(n-1)}\sum_{k=1}^{n}\sum_{l\neq k}|X_{kj}-X_{lj}|\right)^2},\]
\[c_j^{\text{nom}}\!=\!\!\!\sqrt{ \frac1{n(n-1)-1}\sum_{s=1}^{n}\sum_{t\neq s}\!\!\left(\!I\{X_{sj}\neq X_{tj}\}-\frac 1{n(n-1)}\sum_{k=1}^{n}\sum_{l\neq k}\!I\{X_{kj}\neq X_{lj}\}\!\right)^2}.\]
The weights displayed in Figure \ref{fig:MSQWeights} are the minimizing weights for a model with $p=144$, where $X_{i,73},\ldots,X_{i,144}$ are copies of $X_{i,1},\ldots,X_{i,72}$ for all $i=1,\ldots,n$ and $d_1\equiv \ldots\equiv d_{72}\equiv d_{\text{nom}}$ whereas $d_{73}\equiv\ldots \equiv d_{144}\equiv d_{\text{ord}}$. It can be seen that the weights that correspond to $d_{\text{ord}}$ (`ordinal distance') tend to be weighted higher than those that correspond to $d_{\text{nom}}$ (`nominal distance'), which confirms our expectations. Also a binomial test on the signs of the differences $(\omega_{j+72}-\omega_j)$ rejects the null that these differences have median zero with p-value 0.038.
The prediction results shown in Figure \ref{fig:MSQRes} are achieved with $p=72$ and $d_j\equiv d_{\text{ord}}$ for all $j$.

\subsection{Housing in Copenhagen}

\begin{figure}
\begin{centering}
\includegraphics[width=0.8\textwidth]{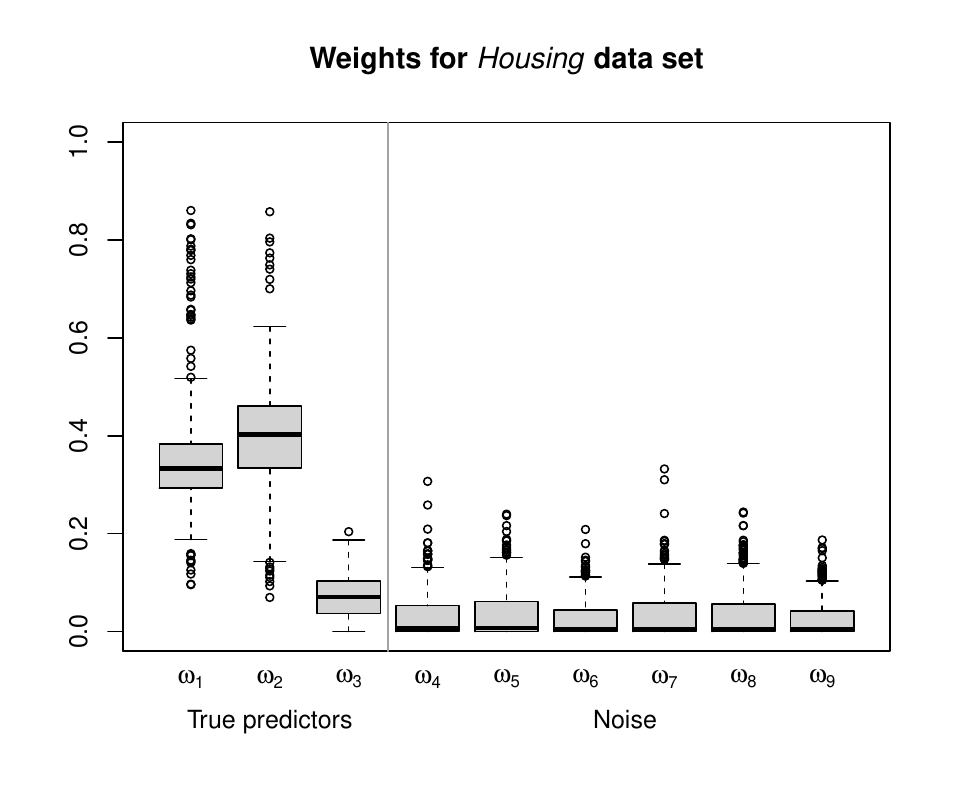}
\caption{Estimated weights (normed) for the {\it Housing} data set with 3 true predictors ($\omega_1,\omega_2,\omega_3$) and 6 additional noise variables ($\omega_4,\ldots,\omega_9$). }\label{fig:HouseWeights}
\end{centering}
\end{figure}

As another example with categorical predictors we consider the Copenhagen housing data with a focus on variable selection. The data set is part of the R-package {\it MASS} by \cite{Rmass}. In the survey 1681 householders in Copenhagen where asked about their satisfaction with their present housing circumstances which could be high, medium or low. We handle this data as a classification problem with $G=3$ and 3 categorical predictors, namely the influence householders have on the management of the property (high, medium or low), the type of rental accommodation (tower, atrium, apartment or terrace), and the contact residents have with other residents (low or high). Additionally, we simulate 6 further categorical covariates that are uniformly distributed on $\{1,2\}$, $\{1,2,3\}$ and $\{1,2,3,4\}$ respectively. Thus our procedure should be able to identify the 3 true predictors in the 9 covariates. In Figure \ref{fig:HouseWeights} the estimated weights are displayed as boxplots over 500 independent repetitions (i.e., simulation of the additional, noise variables has been carried out 500 times). As distance measure we used the ordinal $d_{\text{ord}}$ introduced in the previous example. It can be seen that the 3 true predictors get the highest weights where the third one (contact) seems to have the lowest influence on the satisfaction of the householders.

%The prediction of the class of satisfaction based on a 70/30 split in training and test data works reasonably well for low or high satisfaction level, whereas the level medium seems hard to identify. For comparison we apply the proportional odds method {\it polr} in R-package {\it MASS} on the data as well. The results are very similar both for the model with additional noise as well as for the original data. For level medium the proportional odds model is not able to detect this level at all whereas our method indentifies the level in a few cases. In Figure \ref{fig:HouseRes} we show the prediction results where we apply our procedure on the model with 6 additional noise variables as described above and the proportional odds method on the original data without additional noise.

%
%\begin{figure}
%\begin{centering}
%\includegraphics[width=0.8\textwidth]{MSEHousing.pdf}
%\caption{Prediction results for the {\it Housing} data set based on our procedure for the model with 6 noise variables (blue bars) and on the proportional odds method for the model without additional noise variables (red bars). }\label{fig:HouseRes}
%\end{centering}
%\end{figure}

\section{Concluding Remarks}\label{Conclude}

We proposed a nonparametric method for classification and regression estimation where the covariates may be functional, categorical, or a mixture of both. We allowed for multiple predictors as well as multi-class classification. A key property of our method is its ability of variable/feature weighting, which can also be used for selection purposes.

Although we focussed on functional and categorical predictors, our approach is also suitable for continuous, or continuous mixed with functional and/or categorical, covariates. Due to its universal structure our method works for all types of data that a distance measure can be applied on.

Additionally other loss functions can be considered instead of the Brier Score / the quadratic error. For example in medical applications it could be of interest to minimize false negative results, which is in general also possible with our procedure by adapting the loss function $Q$.

In our extensive simulation study and the application to real world data we showed the good performance of our procedure both in terms of variable weighting/selection as well as estimation and prediction.
An interesting topic in addition could be a thorough theoretical analysis of the asymptotic properties similar to the considerations in \cite{HallLiRacine2007}, who show for a model with continuous and categorical covariates that irrelevant predictors are smoothed out by an optimal bandwidths determination.
However, this is beyond the scope of this paper and will be a topic for future research.

\section*{Statements and Declarations}

There are no relevant financial or non-financial competing interests to report.

\bibliographystyle{spbasic}
\bibliography{BibLeonie}

\end{document}